\documentstyle[12pt,epsf]{article}
\newcommand{\del}{\partial}

\newcommand{\mt}{m_{\mbox{\tiny T}}}
\newcommand{\tauf}{\tau_{\mbox{\tiny F}}}

\newcommand{\Tf}{T_{\!f}}
\newcommand{\vx}{\mbox{\boldmath{$x$}}}
\newcommand{\vv}{\mbox{\boldmath{$v$}}}
\newcommand{\vj}{\mbox{\boldmath{$j$}}}
\newcommand{\vna}{\mbox{\boldmath{$\nabla$}}}

\newcommand{\vp}{\mbox{\boldmath{$p$}}}

\newcommand{\la}{\langle}
\newcommand{\ra}{\rangle}
\begin{document} 
\title{\bf Event-by-event analysis of 
ultra-relativistic heavy-ion collisions 
in smoothed particle hydrodynamics}
\author{
T. Osada${}^{1}$, 
C.E. Aguiar${}^{2}$, 
Y. Hama${}^{1}$
and  
T. Kodama${}^{2}$ 
\\ 
{\it ${}^{1}$Instituto de F\'{\i}sica, 
Universidade de S\~{a}o Paulo,}
\\
{\it ${}^{2}$ 
Instituto de F\'{\i}sica, 
Universidade Federal do Rio de Janeiro}
}
\date{\today}
\maketitle
\begin{abstract}
The method of smoothed particle hydrodynamics (SPH) is applied 
for ultra-relativistic heavy-ion collisions. The SPH method has 
several advantages in studying event-by-event fluctuations, which 
attract much attention in looking for quark gluon plasma (QGP) 
formation, because it gives a rather simple scheme for solving 
hydrodynamical equations. 
Using initial conditions for Au+Au collisions at RHIC energy 
produced by NeXus event generator, we solve the hydrodynamical 
equation in event-by-event basis and study the fluctuations of 
hadronic observables such as $dN/dy$ due to the initial conditions. 
In particular, 
fluctuations of elliptic flow coefficient $v_2$ is 
investigated for both the cases, with and without QGP formation. 
This can be used as an additional test of QGP formation. 
\end{abstract}
\section{Introduction} 
One of the central issues of the high-energy heavy-ion physics is to 
investigate the phases of the hadronic and quark matter.
Hydrodynamical descriptions\cite{Landau}, which have a rather long 
history, may be a powerful tool for studying such phase diagrams 
because we can handle the equation of state directly in the model. 
However, to extract precisely information about the equation of state,
we must be careful in comparing the model predictions with experimental 
data because procedures of event averaging may make 
signal of QGP ambiguous. 

Event-by-event analysis\cite{Heiselberg00} is one of promising ways to 
extract clear information from the experimental data, in particular at 
RHIC and LHC energy regions. The fluctuations at the initial stage 
(for example, the initial energy density fluctuation) must 
affect the formation of QGP and the later space-time 
evolution of the 
whole system. Fig.1 shows the energy density (counter plot, at 
$z=0~fm$ plane) for a Au+Au collision at $\sqrt{s} =200A~GeV$, 
$b=0~fm$, produced by 
NeXus code\cite{NeXus}. As shown in Fig.1, the fluctuations at the 
initial stage of relativistic heavy-ion collisions are not 
negligible. 
To achieve an event-by-event analysis using relativistic hydrodynamical 
model, the numerical code must deal with arbitrary initial conditions 
and equation of state with suitable calculational speed and
precision. The method of smoothed particle
hydrodynamics (SPH)\cite{SPH-nu}, which was first introduced for 
astrophysical applications\cite{SPH-as}, satisfy such requirements. 
The main characteristic of SPH is the introduction of 
``particles'' attached to some conserved quantity 
which are described in terms of discrete Lagrangian coordinates. 
This feature of the model enables to easily carry out calculations of 
the ultra-relativistic heavy-ion collisions, which is accompanied by 
prominent longitudinal expansion. 
Another advantage of using SPH method is that we can choose a suitable
precision in solving hydrodynamical equations because of the smoothing
kernel, which allows to smooth out often unnecessary very precise
local aspects and then achieve a reduction of computational time. 
This is a very profitable point in the event-by-event analysis. 

In the following section, we briefly formulate 
the entropy-based SPH, using the variational approach\cite{Elze99}. 
In Sec.3, to check the performance of SPH, we apply it for solving 
problems of relativistic hydrodynamics whose solutions are known 
analytically or numerically. 
Then, in Sec.4, we demonstrate hydrodynamical evolutions 
of both resonance gas and quark gluon plasma 
starting from the same simple initial conditions. 
In Sec.5, after giving a brief explanation how to take 
the initial conditions of NeXus into our code, we present typical 
hydrodynamical evolution of an event at RHIC energy. 
To introduce freeze-out process into our code, we formulate 
Cooper-Frye formula in terms of SPH and show several results 
on single-particle spectrum in Sec.6. 
Then, we study the event-by-event fluctuation of elliptic flow coefficient 
$v_2$\cite{Voloshin96} at RHIC energy\cite{STAR00} in Sec.7, showing 
that the dispersion is large and 
it does depend on the equation of state used. 
Finally we close our discussions with some concluding remarks. 

\section{Relativistic SPH equations} 
\subsection{SPH representation} 
To specify the hydrodynamical state of a fluid, we need to 
locally know the thermodynamical quantities such as the energy 
density $\varepsilon$ and the collective velocity field, 
besides equation of state 
of the fluid. If local thermal equilibrium is well established, 
the thermodynamical variables may be expressed by 
smooth functions of the coordinates. Let us choose $N$ 
space points to specify the state of the whole system. 
To know the physical quantities in other space points, we 
may use an interpolation using the known $N$ points. 
(Hereafter we use the word `particles' instead of `points'.) 

In SPH, for an arbitrary extensive thermodynamical quantity of 
$A$, the density (in the space-fixed frame) $a^*$ is 
parametrized in the following way:
\begin{eqnarray}
    a^*(\vx,t)= \sum_i^N \nu_i~ W(\vx-\vx_i(t);h)~, 
    \label{SPH-repre}  
\end{eqnarray} 
where $W(\vx-\vx_i(t);h)$ is a positive definite kernel function 
with the properties 
\begin{eqnarray} 
 \int \! d\vx'~W(\vx'-\vx) \equiv 1 
 \quad\quad \mbox{and}
 \quad\quad   W(\vx'-\vx,h) = 
 \delta(\vx'-\vx) \quad \mbox{when $h\to 0$} ~, 
 \nonumber 
\end{eqnarray} 
where $h$ is a smoothing-scale parameter. 
Using the normalization above, we have 
\begin{eqnarray} 
  A_{total} = \int\! d^3\vx~a^*(\vx,t)= \sum_i^N \nu_i ~. 
  \label{A_total}
\end{eqnarray} 
From eq.(\ref{A_total}) the quantity $\nu_i$ can be interpreted 
as a portion of $A$ which is carried by 
the $i$th particle. 
Its velocity, identified as the velocity of the 
fluid at $\vx_i(t)$, satisfies  
\begin{eqnarray}
 \vv_i(t) = \frac{d\vx_i(t)}{dt}~. 
 \label{velocity} 
\end{eqnarray}
If $A$ is a conserved quantity, $\nu_i$ must be constant in time. 
In this case, from eqs.(\ref{SPH-repre}) and (\ref{velocity}), 
we obtain 
\begin{eqnarray}
    \frac{\del }{\del t} ~a^*(\vx,t) = -\vna\cdot\vj(\vx,t)~,
    \label{continuity}
\end{eqnarray} 
where 
\begin{eqnarray}
     \vj(\vx,t) = \sum_i ~\nu_i \vv_i(t) ~W(\vx-\vx_i(t);h) 
\end{eqnarray} 
is interpreted as the current density of the thermodynamical quantity 
$A$ and eq.(\ref{continuity}) is then the corresponding continuity equation.  

\subsection{Equations of motion for SPH particles} 
To know the time evolution of the fluid, we have to find the equation 
of motion for each SPH particle. It can be obtained by the 
variational method\cite{Elze99} with the following Lagrangian\cite{SPH-nu} 
\begin{eqnarray}
  L_{\mbox{\tiny SPH}} (\{ \vx_i,\vv_i\} ) = 
  - \sum_i \nu_i ~\frac{\varepsilon_i}{a^*_i} 
  \label{Lagrangian1} 
\end{eqnarray} 
It should be noted here that the factor $\nu_i/a^*_i$ can be 
regarded as the volume $V_i$ (in the space-fixed frame) occupied 
by the $i$th `particle'. Then each term in eq.(\ref{Lagrangian1}) 
is equal to $E_i/\gamma_i$, where $E_i$ and $\gamma_i$ are 
the energy (in the rest frame) and the gamma factor of the 
$i$th particle, respectively. 
Then we also write 
\begin{eqnarray}
   L_{\mbox{\tiny SPH}} (\{ \vx_i,\vv_i\} )
   = - \sum_i \frac{E_i(\vx_i,\vv_i)}{\gamma_i(\vv_i)} ~. 
  \label{Lagrangian2} 
\end{eqnarray} 

The equations of motion of SPH particles 
are obtained minimizing the action $I_{\mbox{\tiny SPH}}$ 
with respect to $\{\vx_i\}$ with the constraint given by 
eq.(\ref{velocity}) taken into account 
\begin{eqnarray}
   \delta 
    I_{\mbox{\tiny SPH}} (\{ \vx_i,\vv_i\} ) = 
   -\delta \int\!\!dt~~ \frac{E_i}{\gamma_i} =0~.
   \label{mini-I} 
\end{eqnarray} 
Using the adiabatic condition 
\begin{eqnarray}
     \delta E_i = P\delta V_i \quad = 
     P_i ~\delta\Big( \frac{\nu_i\gamma_i}{a^*_i}\Big)~, 
\end{eqnarray} 
we obtain 
\begin{eqnarray}
      \frac{d}{dt}(\nu_i\frac{P_i+\varepsilon_i}{a_i}\gamma_i\vv_i)
    + \sum_{j}\nu_j
      \bigg[ 
      \frac{P_i}{{a^*_i}^2} + 
      \frac{P_j}{{a^*_j}^2}\bigg] 
     ~\vna W(\vx_i-\vx_j;h) =0 ~, 
    \label{eq.motion}
\end{eqnarray} 
where $P_i$ is the pressure. Because of the adiabatic condition, 
eqs.(\ref{eq.motion}) are the equations of motion for the ideal fluid. 
For the extension to the non-adiabatic case, see ref.\cite{SPH-nu}.  

\subsection{Entropy based SPH in general coordinate system} 
As a possible candidate for the thermodynamical extensive quantity 
$a^*$, we may consider baryon number density, entropy density, 
energy density and so on. In the case of energy density, 
we have
(cf. eq.(\ref{Lagrangian1}) and (\ref{Lagrangian2})) 
\begin{eqnarray}
   \nu_i = E_i(\vx_i,\vv_i) ~. 
\end{eqnarray} 
However, this is only the internal energy, so $\nu_i$ is not 
constant with respect to the variation of ${\vx_i}$ 
in eq.(\ref{mini-I}), introducing an additional 
complication without any practical merit. 
On the other hand, for entropy or baryon number density, for example, 
$\nu_i$ can be kept constant. 
Since we will apply SPH to ultra-relativistic heavy-ion collisions, 
baryon number density is not so suitable choice because it is 
expected to be very small in the central rapidity region. 
Let us choose the entropy density for $a^*$ in eq.(\ref{SPH-repre}).

In relativistic 
heavy-ion collisions, initial conditions of fluid are often 
given at constant $\tau\equiv\sqrt{t^2-z^2}$. 
Then, it is convenient to formulate SPH in the coordinates 
\begin{eqnarray} 
    x^0\equiv\tau = \sqrt{t^2-z^2},~~
    x^1\equiv x,~~
    x^2\equiv y,~~
    x^3\equiv\eta = \frac{1}{2}\ln\frac{t+z}{t-z}~~. 
    \label{coordinate}
\end{eqnarray}
The SPH equations in generalized coordinate system are easily 
derived in a similar way\cite{SPH-nu}. 
In the case of the coordinate above, the SPH equations of 
motion should be written as 
\begin{eqnarray}
  && s(\tau,\vx) = 
  \frac{1}{\tau u^0} \sum_i\nu_i W(\vx-\vx_i:h) ~,
  \label{sph-repre-s}\\ 
  && \frac{d}{d\tau}(\nu_i\frac{P_i+\varepsilon_i}{s_i}\gamma_i\vv_i)
    + \sum_{j}\frac{\nu_j}{\tau} 
      \bigg[ 
      \frac{P_i}{{s_i}^2{u^0_i}^2} + 
      \frac{P_j}{{s_j}^2{u^0_j}^2}\bigg] 
     ~\vna W[\vx_i-\vx_j;h] =0  ~. 
\end{eqnarray} 
\section{Numerical check of SPH} 
To check the performance of SPH formulated in Sec.2, 
we applied it to a problem of relativistic hydrodynamics 
whose solution is known analytically. 
When the longitudinal rapidity $\alpha$ 
and the transverse one $\beta$ are given as 
\begin{eqnarray}
  \alpha \equiv  \eta,\quad 
  \beta  \equiv  
  \frac{1}{2}\ln \frac{\tau+\sqrt{x^2+y^2}}{\tau-\sqrt{x^2+y^2}}~, 
\end{eqnarray}
it is known that the entropy density is given by 
\begin{eqnarray}
  s= \frac{s_0}{[\tau^2-x^2-y^2]^{3/2}}~,
  \label{3dscaling}  
\end{eqnarray} where $s_0$ is a
constant which is determined by the initial conditions. 
A comparison of the numerical results by SPH with the 
analytical solution eq.(\ref{3dscaling}) is 
shown in Fig.2, where the equation of state used was $c_s^2=1/3$. 
It is verified that SPH solves this problem correctly. 
For a further check, we applied it also to another problem, 
namely, relativistic hydrodynamics with the longitudinal scaling 
and the Landau-type initial condition in the transverse directions, 
\begin{eqnarray}
  \alpha \equiv \eta, \quad 
  \beta  \equiv 0 ~. 
\end{eqnarray}
Figure 3 shows a comparison of the SPH results with those 
obtained by Hama and Pottag\cite{Pottag84}. 
The initial conditions and the equation of state used 
are exactly the same as in ref.\cite{Pottag84}. 
We found that SPH works well also in this problem. 

\section{Equations of state and hydrodynamical evolutions} 
To complete the theory of relativistic hydrodynamics, 
the equation of state (EoS) for the fluid is required. 
We use two possible EoS in the present study, namely, a 
resonance gas EoS\cite{Shuryak72} and 
an EoS containing both resonance gas and QGP via a first order 
phase transition\cite{Hung95}. The EoS are parametrized 
as follows (See also Fig.4.):  
\begin{eqnarray}
   P/\varepsilon = 0.20 \mbox{\quad ( constant for whole
   $\varepsilon$~region ), }
\end{eqnarray} 
for the resonance gas (RG) and 
\begin{eqnarray}
   P/\varepsilon =\left\{
   \begin{array}{ll}
   0.20                & 
   \quad\quad 0.00 ~GeV/fm^3~<~ \varepsilon ~<~ 0.28 ~GeV/fm^3 
   \equiv \varepsilon_1 \\
   0.056/\varepsilon   & 
   \quad\quad 0.28 ~GeV/fm^3 ~<~ \varepsilon ~<~ 1.45 ~GeV/fm^3 
   \equiv \varepsilon_2 \\
   1/3-4B/3\varepsilon & 
   \quad\quad 1.45 ~GeV/fm^3 ~<~ \varepsilon 
   \end{array}\right. 
\end{eqnarray} 
for the quark gluon plasma (QGP) EoS. The bag constant $B=0.32 GeV/fm^3$ 
is determined to give the critical temperature $T_c=160~MeV$. 
A remark should be made here that when we call QGP EoS, here and 
in the following sections, we are actually considering all these three 
states according to the value of $\varepsilon$. 

We show some results of the hydrodynamical evolution 
using such RG and QGP EoS in Figs. 5 and 6. 
For the both cases, the initial ($\tau_0 =1.0~fm$) entropy density $s$, 
longitudinal and transverse rapidity $\alpha$, $\beta$ 
are set as 
\begin{eqnarray}
    s= 40 [1/fm^3], \quad \alpha = \eta~,\quad \beta = 0 ~.
    \quad \mbox{for  $\sqrt{x^2+y^2}<7.0~fm$,  $|\eta|<7.0$ ~.} 
\end{eqnarray} 
As seen in Figs.5 and 6, the space-time evolution of the fluid 
strongly depends on the EoS. 
In comparison with the RG case, the temperature of the fluid in 
QGP EoS drops much slowly due to the existence of the mixed phase, 
where the pressure is constant. 

\section{Initial conditions produced by NeXus} 
The most fundamental assumption of hydrodynamical models is 
the thermalization of partonic or hadronic systems produced after 
high-energy heavy-ion collisions. 
In our study, we set the initial 
conditions using information from Nexus event 
generator\cite{NeXus}, 
in the following 2 steps. 
\begin{enumerate}
  \item Align all primary hadrons on the $\tau=\tau_0$ hypersurface. 
Here we assume that the primary hadrons move freely with the 
momentum $p^{\mu}$. In the present study, we use $\tau_0=1.0~fm$. 
  \item The initial energy density and momentum density are estimated 
using the interpolation kernel $W$. 
  \begin{eqnarray*}
 &&\varepsilon(\tau_0;\vx)= \sum_i  p^{0}_i W(\tau_0,\vx-\vx_i;h) ~,\\
 &&\vp(\tau_0;\vx)=         \sum_i \vp_i W(\tau_0,\vx-\vx_i;h)   ~,
  \end{eqnarray*} 
where $\vx_i=(x_i,y_i,\eta_i)$ are the coordinates of the 
$i$-th primary hadron after the alignment on $\tau=\tau_0$. 
The smoothing scale parameter $h=1~fm$ is used, which is 
the rough size of the hadrons. The collective velocity field is then 
given by 
$\vv(\tau_0,\vx)= \vp(\tau_0,\vx)/\varepsilon(\tau_0,\vx)$~. 
  \end{enumerate}
In Fig.7, we show the hydrodynamical evolution of a typical 
event of $Au+Au$ collision at energy $\sqrt{s}=130A~GeV$ 
for the impact parameter $b=7.0fm$ given by our hydro code, 
SPheRIO\footnote{
{\bf S}moothed {\bf P}article {\bf h}ydrodynamical {\bf e}volution  
of {\bf R}elativistic heavy {\bf IO}n collisions}, connected to the 
NeXus event generator. 

\section{Cooper-Frye formula in SPheRIO} 
The Cooper-Frye formula\cite{Cooper74} is widely used 
in calculations of single particle spectra using 
hydrodynamical models because of its simplicity. 
So, as the first trial to evaluating single 
particle densities, we shall follow it. 

In SPH, using eq.(\ref{sph-repre-s}), the usual 
Cooper-Frye formula can be rewritten as 
\begin{eqnarray}
  E\frac{d^3N}{d\vp^3}
  \!\!&=&\!\! 
  \frac{g}{(2\pi)^3}
  \sum_j \nu_j \int_{F.O.}\hspace*{-3mm}
  \tau dxdyd\eta~ 
  \frac{
  \sigma_{\mu}p^{\mu}~  
  }{\tau ~su^0}~  
  W(\vx-\vx_j(\tau);h)~
  \frac{1}{e^{u_{\mu}p^{\mu}/\Tf}\mp 1}~,     
  \label{CF1} 
  \end{eqnarray} 
where $\sigma_{\mu}$ is defined by 
\begin{eqnarray}
   \sigma_{\mu}\equiv \Big(~1,\frac{n_{x}}{n_{\tau}},
   \frac{n_{y}}{n_{\tau}}, \frac{n_{\eta}}{n_{\tau}}~\Big) 
  \end{eqnarray} 
and $n_{\mu}=(n_{\tau},n_x,n_y,n_{\eta})$ is the normal vector 
of the isothermals: 
\begin{eqnarray}
    n_{\mu} \equiv  \Big( -\frac{\del T}{\del\tau},-\frac{\del T}{\del x},
                         -\frac{\del T}{\del y},-\frac{\del
         T}{\del\eta}~\Big) ~. 
  \end{eqnarray}
The integration is usually done over a constant temperature 
hypersurface, $T=T_f$. 
When the spline-type kernel function\cite{SPH-nu,SPH-as} is used, 
$W(\vx-\vx';~h)=0$ for  $|\vx-\vx'|>2h$. Then, within the approximation 
of small $h$, 
\begin{eqnarray} 
  E\frac{d^3N}{d\vp^3}
  \!\!&\approx&\!\!   
  \frac{g}{(2\pi)^3} 
  \sum_j \nu_j 
  \bigg[ 
  \frac{p^{\mu}}{e^{u_{\mu}p^{\mu}/\Tf}\mp 1} \bigg]_j 
  \int
  \tau dxdyd\eta~ 
  \frac{
  \sigma_{\mu}~  
  }{\tau ~su^0}~  
  W(\vx-\vx_j(\tauf(\vx))~;h)~\nonumber \\
  \!\!&=&\!\! 
  \frac{g}{(2\pi)^3} 
  \sum_j \bigg[ 
  \frac{p^{\mu}}{e^{u_{\mu}p^{\mu}/\Tf}\mp 1} \bigg]_j 
  \int_{}~ 
  dxdyd\eta~   \frac{
  \nu_j 
  }{~su^0}~
  \frac{\sigma_{\mu} W(\vx-\vx_j~;h)}{|1+v_j^{x}\frac{n_{x}}{n_{\tau}}
          +v_j^{y}\frac{n_{y}}{n_{\tau}}
          +v_j^{\eta}\frac{n_{\eta}}{n_{\tau}}|}~, 
  \label{CF2} 
  \end{eqnarray} 
where $\vx_j$ gives the point where $j$-th SPH-particle 
crosses the freeze-out hypersurface. 
The result of the single-particle spectra, 
$1/\mt dN/d\mt$, $dN/dy$ and $dN/d\varphi$ for Au+Au collisions 
at $\sqrt{s}=130A~GeV$, $b=7.0~fm$ are shown 
in Figs. 8, 9 and 10, respectively. The freeze-out 
temperature used is $T_f=140~MeV$. 

\section{Event-by-event analysis of hadronic observables} 

\subsection{Elliptic flow coefficient $v_2$} 
The flow phenomena\cite{Danielewicz00} can be important candidates 
of QGP signals because they may carry much information about EoS 
during the expansion of quark or hadronic matter. In particular, 
the elliptic flow coefficients\cite{Voloshin96} is one of 
the interesting observables 
since it is sensitive to the early stage of high-energy heavy-ion 
collisions. Because it is expected that different EoS responds to the 
initial fluctuations in different way, the measure of fluctuation, 
for example, $\delta v_2 \equiv \sqrt{\la v_2^2\ra-\la v_2\ra^2}$ is 
also very important. For this purpose, we investigate $v_2$ 
distribution using different EoS. In Figs.11, 12 and Table 1, 
we show results of the $v_2$ distribution for $Au+Au$ collisions 
at energy $\sqrt{s}=130A~GeV$ in RG and QGP EoS cases. 

As seen in Table 1, the fluctuations of flow coefficient 
$v_2$ in QGP is about 20 $\sim$ 30\% smaller than that 
in RG (cf. the difference of $\la v_2 \ra$ in RG and QGP. 
It is about a few \%.). The $b$ dependence of 
$\delta v_2$ is also found. 
Hence, it is expected that $\delta v_2$ also brings us useful 
information to infer on the phase passed 
in the early stage of high-energy nuclear collisions. 

\subsection{Fluctuation of the slope parameter} 
It is also interesting to study the fluctuation of the 
so-called slope parameter in the transverse momentum spectra. 
The slope parameter $\tilde{T}$ can be obtained by fitting 
${1}/{\mt}{dN}/{d\mt}$ with a function ${\cal N}
\exp{[-\mt/\tilde{T}]}$. In the present study, 
$0<\mt-m_{\pi}<1.0~GeV$ interval is used for the fitting. The results 
are shown in Table 2. We found $\tilde{T}$ and $\delta \tilde{T}$ 
of QGP is systematically smaller than that of RG. 

\subsection{Multiplicity fluctuation in the central region} 
Other interesting measurement may be multiplicity fluctuation. 
Let us define the fluctuation measure as 
\begin{eqnarray}
     \delta n(y,\Delta y) \equiv 
     \sqrt{ \la n^2(y,\Delta y) \ra -\la n(y,\Delta y)\ra^2 }
  \end{eqnarray} 
where 
\begin{eqnarray}
    n(y,\Delta y) = 
    \int_{y-\Delta y}^{y+\Delta y} \!\!\!\! dy~\frac{dN}{dy}~. 
  \end{eqnarray} 
We investigate the EoS dependence of the multiplicity fluctuation 
in two central ($y=0$) rapidity windows $\Delta y$ and as function of the 
impact parameter $b$. The results are summarized in Table 3. 
As seen there, the multiplicities and their 
fluctuations $\delta n$ have a 
clear EoS dependence. See Figs. 13 and 14. 
When the initial energy density is bellow $\varepsilon_2$ 
the multiplicity fluctuation with QGP EoS  
is expected to be larger than that of RG without 
first order phase transition. 
This is due to the existence of the mixed phase (see Appendix). 
The values of ratio $\delta n/\la n\ra$ of QGP is slightly 
larger than those of RG. 

\section{Concluding remarks and discussions} 
We developed a new hydrodynamical code, SPheRIO, based on the 
smoothed particle hydrodynamics (SPH), for studying high-energy 
heavy-ion collisions in event-by-event basis. In this study, 
we set the initial conditions based on the results of NeXus 
event generator and used two types of possible equation of 
states (EoS), $ie.$, the resonance gas (RG) and quark gluon 
plasma (QGP) EoS. In the freeze-out process, single 
particle spectra are calculated by the Cooper-Frye formula for 
each set of initial conditions 
and EoS. Analyzing these single particle spectra, 
the fluctuations in flow coefficients, slope parameter 
$\tilde{T}$ and multiplicities in the central rapidity 
regions are investigated at RICH energy. 

As seen in the results of Table 1, we found remarkable 
difference in $\delta v_2$ between 
RG and QGP EoS predictions. This means that the measure of the 
fluctuations is also helpful to infer the phase passed 
in the high-energy heavy-ion collisions. 
In the QGP EoS case, the fluctuations are about $20\sim30$ \% 
smaller than those in RG case. Our predictions of the 
average value of $v_2$ is shown in Fig.15, compared with 
STAR Collaboration data\cite{STAR00}. 
It is seen that the main trend of the data is reproduced, 
but the fluctuation is quite large.  

The fluctuation of the slope parameter $\tilde{T}$ is small, 
a few \% of $\tilde{T}$. We found the fluctuation of multiplicity 
has a clear EoS dependence. The impact parameter dependence of 
$\delta n(y,\Delta y)$ can be a good measure for the detection of 
the first order phase transition. 

\vspace*{5mm} 

\noindent 
{\bf Acknowledgments:} 
The authors acknowledge stimulating discussions with 
K. Werner, O. Socolwski Jr and H.J. Drescher. 
This work was partially supported 
by FAPESP(contract no.s 98/02249-4 and 98/00317-2), 
PRONEX(contract no. 41.96.9886.00), FAPERJ(contract
no. E-26/150.942/99) and CNPq. 

\appendix 
\renewcommand{\theequation}{\Alph{section}\arabic{equation}}
\newcommand{\myeqnarray}[1]{\section{#1}\setcounter{equation}{1}}
\setcounter{equation}{0} 
\section{Entropy fluctuations in the first order phase transition}
\subsection{Mixed phase dominant case:} 
If in the most part of the fluid the energy density at the 
initial time $\tau_0$ varies only in the interval 
$\varepsilon_1 < \bar{\varepsilon}(x) <\varepsilon_2$, 
where $\varepsilon_1$ and $\varepsilon_2$ are the critical densities 
of the RG and QGP phases, respectively, the produced entropy $S$ of 
the system is 
\begin{eqnarray}
   S_{} \!\!&=&\!\! 
       \int \! d^3 x ~s_1~\Big[~
       \frac{\bar{\varepsilon}(x)+\delta\varepsilon(x) 
       }{\varepsilon_1}~\Big]^{1/[1+c_{m}^2]} 
       \approx 
        \int \! d^3 x~s_1
        \Big(\frac{\bar{\varepsilon}(x)}{\varepsilon_1}
        \Big)^{}~\Big[~1
        +\Big(\frac{\delta\varepsilon(x)}{\bar{\varepsilon}(x)}\Big)
        ~\Big] \nonumber \\
  \end{eqnarray} 
where $c_{m}=0$(constant) 
is the sound velocity in the mixed phase(Mix) and 
$s_1$ is the critical entropy density corresponding to
$\varepsilon_1$. 
$\bar{\varepsilon}(x)$ and $\delta{\varepsilon}(x)$ 
denote the event averaged energy density and fluctuation, 
respectively. Here, we assume that 
$\delta\varepsilon(x)\ll\bar{\varepsilon}(x)$, so that 
the mixed phase is dominant in all the events. 
\noindent 
Then, the dispersion of the total 
entropy $S$ is 
\begin{eqnarray}
    D^2_{Mix} = \la S^2 \ra -\la S \ra^2
    \approx 
    \int\! d^3 x~~
    s_1^2~\Big(\frac{\bar{\varepsilon}(x)}{\varepsilon_1}
    \Big)^{2}  
    ~\bigg\la\Big(
    \frac{\delta\varepsilon(x)}{\bar{\varepsilon}(x)}\Big)^2\bigg\ra 
\end{eqnarray} 
In the case the phase transition is absent, so that the 
fluid is described by RG EoS, the same events with the same 
energy distribution and fluctuation gives 
\begin{eqnarray}
   S_{RG} \!\!&=&\!\! 
       \int\! d^3 x ~s_1~\Big[~
       \frac{\bar{\varepsilon}(x)+\delta\varepsilon(x) 
       }{\varepsilon_1}~\Big]^{1/[1+c_{r}^2]} \nonumber \\
       \!\!&\approx&\!\! 
        \int\! d^3 x~s_1
        \Big(\frac{\bar{\varepsilon}(x)}{\varepsilon_1}
        \Big)^{5/6}
        ~\Big[~1+\frac{5}{6}
        \Big(\frac{\delta\varepsilon(x)}{\bar{\varepsilon}(x)}\Big)
        -\frac{5}{72}
        \Big(\frac{\delta\varepsilon(x)}{\bar{\varepsilon}(x)}\Big)^2
        +\cdots~\Big]
  \end{eqnarray} 
and 
\begin{eqnarray}
    D_{RG}^2 = \int\! d^3 x~~
    s_1^2~\Big(\frac{\bar{\varepsilon}(x)}{\varepsilon_1}
    \Big)^{5/3}  ~\frac{25}{36}
    \bigg\la\Big(
    \frac{\delta\varepsilon(x)}{\bar{\varepsilon}(x)}\Big)^2\bigg\ra
    ~, 
  \end{eqnarray} 
respectively. We used the sound velocity 
$c_r=1/\sqrt{5}$(constant) here. 
So, the appearance of the mixed phase produces 
\begin{eqnarray*} 
   \Delta D^2\!\!&\equiv& 
   D^2_{Mix}-D^2_{RG}=~
   \!\!\int\! d^3x~s_1^2\bigg\{~\Big(
   \frac{\bar{\varepsilon}(x)}{\varepsilon_1}\Big)^2 -
   \frac{25}{36}
   \Big(\frac{\bar{\varepsilon}(x)}{\varepsilon_1}\Big)^{5/3}
   ~\bigg\}
   \bigg\la\Big(
   \frac{\delta\varepsilon(x)}{\bar{\varepsilon}(x)} \Big)^2\bigg\ra 
   ~~>0 
   \end{eqnarray*} 
Hence, 
\begin{eqnarray}
  D_{Mix}^2 > D^2_{RG} ~. 
  \end{eqnarray} 
\subsection{QGP phase dominant case:}
If the most part of the fluid is in the QGP phase at initial time 
$\tau_0$, the total produced entropy $S$ of the system is
\begin{eqnarray}
   S_{QGP} \!\!&=&\!\! 
       \int\!  d^3 x ~s_2~\Big[~
       \frac{\bar{\varepsilon}(x)+\delta\varepsilon(x) 
       }{\varepsilon_2}~\Big]^{1/[1+c_{q}^2]} \nonumber \\
       \!\!&\approx&\!\! 
        \int\!  d^3 x~s_1
        \Big(\frac{\varepsilon_2}{\varepsilon_1}\Big)~
        \Big(\frac{\bar{\varepsilon}(x)}{\varepsilon_2}
        \Big)^{3/4}
        ~\Big[~1+\frac{3}{4}
        \Big(\frac{\delta\varepsilon(x)}{\bar{\varepsilon}(x)}\Big)
        -\frac{3}{32}
        \Big(\frac{\delta\varepsilon(x)}{\bar{\varepsilon}(x)}\Big)^2
        +\cdots~\Big]
  \end{eqnarray}
and 
\begin{eqnarray}
    D_{QGP}^2 = \int\!  d^3 x~~
    s_1^2
    \Big(\frac{\varepsilon_2}{\varepsilon_1}\Big)^2~
    ~\Big(\frac{\bar{\varepsilon}(x)}{\varepsilon_2}
    \Big)^{3/2}  ~\frac{9}{16}
    \bigg\la\Big(
    \frac{\delta\varepsilon(x)}{\bar{\varepsilon}(x)}\Big)^2\bigg\ra ~,
  \end{eqnarray}
where $c_q=1/\sqrt{3}$(constant) is assumed. 
Then, we have 
\begin{eqnarray} 
   \Delta D^2\!\!&\approx& 
   D^2_{QGP}-D^2_{RG}=~
   \!\!\int\! d^3x~s_1^2\bigg\{~\frac{9}{16}
   \Big(\frac{\varepsilon_2}{\varepsilon_1}\Big)^2~
   \Big(
   \frac{\bar{\varepsilon}(x)}{\varepsilon_2}\Big)^{3/2} 
   -
   \frac{25}{36}
   \Big(\frac{\bar{\varepsilon}(x)}{\varepsilon_1}\Big)^{5/3}
   ~\bigg\}
   \bigg\la\Big(
   \frac{\delta\varepsilon(x)}{\bar{\varepsilon}(x)}
   \Big)^2\bigg\ra~. 
   \nonumber \\ 
  \end{eqnarray}  
From this relation, we can conclude that when the energy density 
$\bar{\varepsilon}(x)$ becomes larger than a certain value, 
$(0.81)^6~(\varepsilon_2/\varepsilon_1)^3\varepsilon_1~~
\approx 11~GeV/fm^3 $~~for the most part of the fluid, 
the dispersion of the QGP case becomes smaller than RG case. 

%
\begin{figure}[tbh]
{\vspace*{-0.0cm} {\hspace*{+0.0cm} \epsfysize=14cm %
\epsfbox{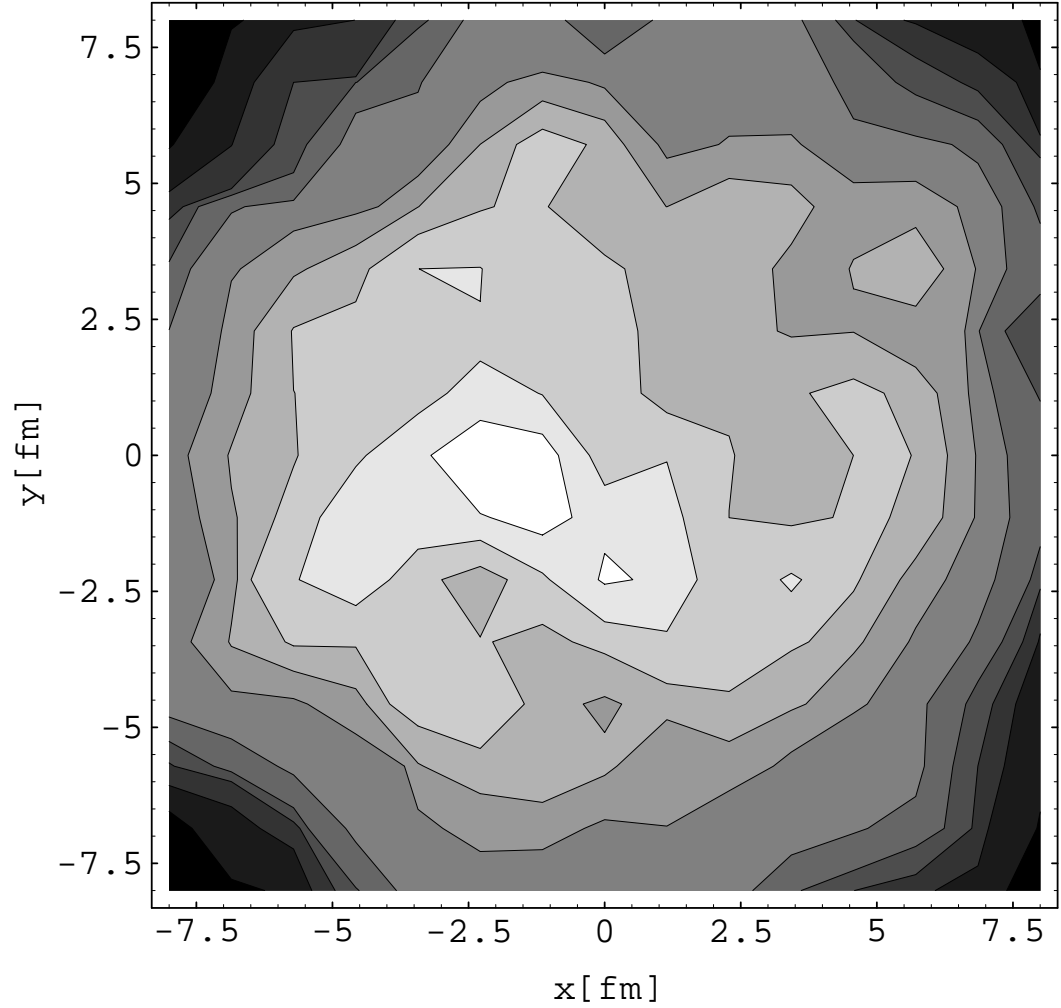}}}
\caption{Initial energy density (counter plot $z=0~fm$ plane) 
of a typical event of Au+Au collision at energy $\sqrt{s}=200A~GeV$, 
impact parameter $b=0 fm$, produced by NeXus event generator\cite{NeXus}.} 
\vspace*{0.0cm} 
\end{figure}
\clearpage 
\begin{figure}[tbh]
{\vspace*{-0.7cm} {\hspace*{1.5cm} \epsfysize=10.0cm %
\epsfbox{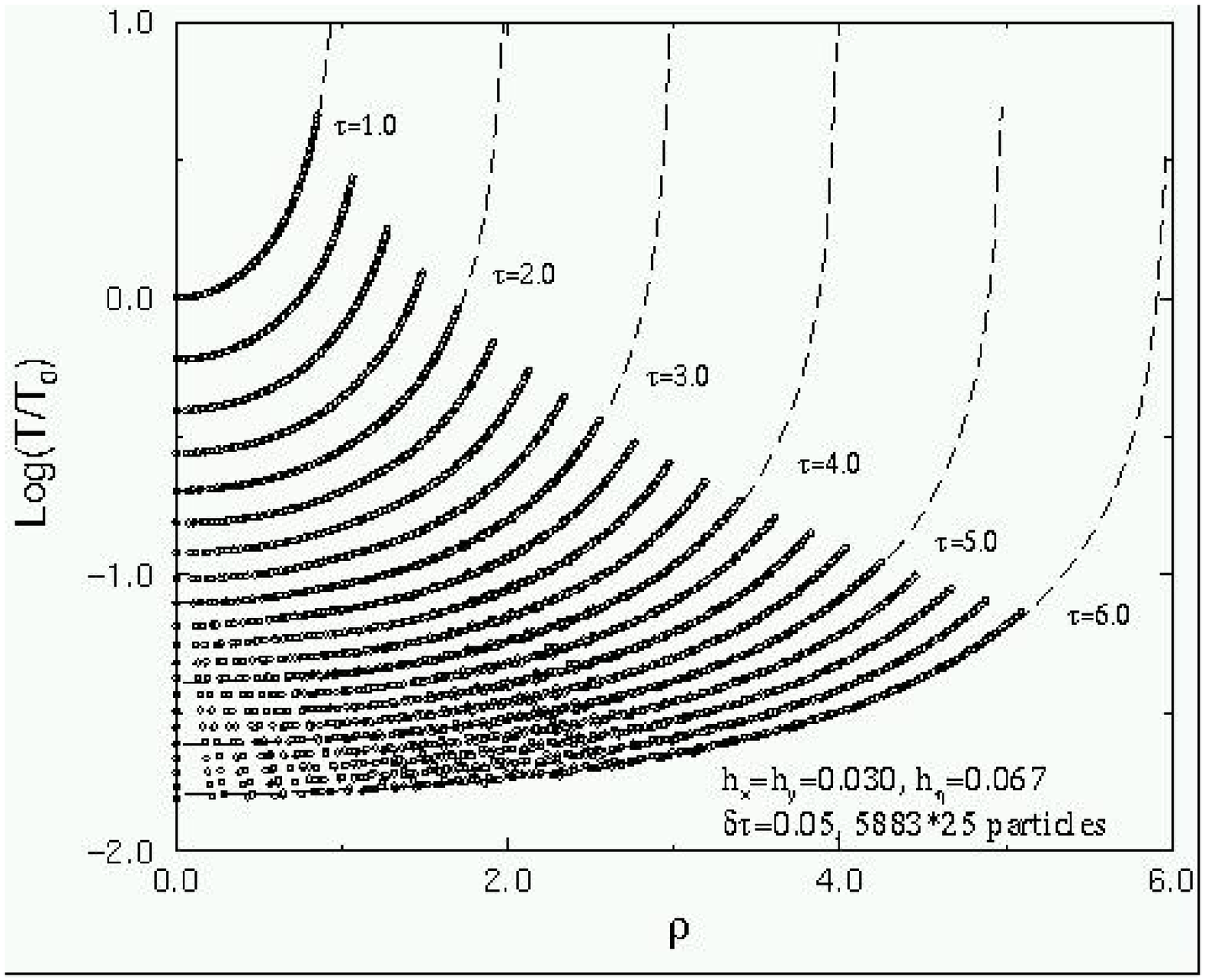}}
\vspace*{0.0cm} 
\caption{Comparison of results by SPH (open circles) 
with the analytical solution, eq.(\ref{3dscaling}).
The sizes of the parameters $h_x(=h_y)$, $h_{\eta}$ and 
$d\tau$ used were 0.030, 0.067 and 0.05, 
respectively. 
}}
{\vspace*{0.5cm} {\hspace*{1.5cm} \epsfysize=10.0cm %
\epsfbox{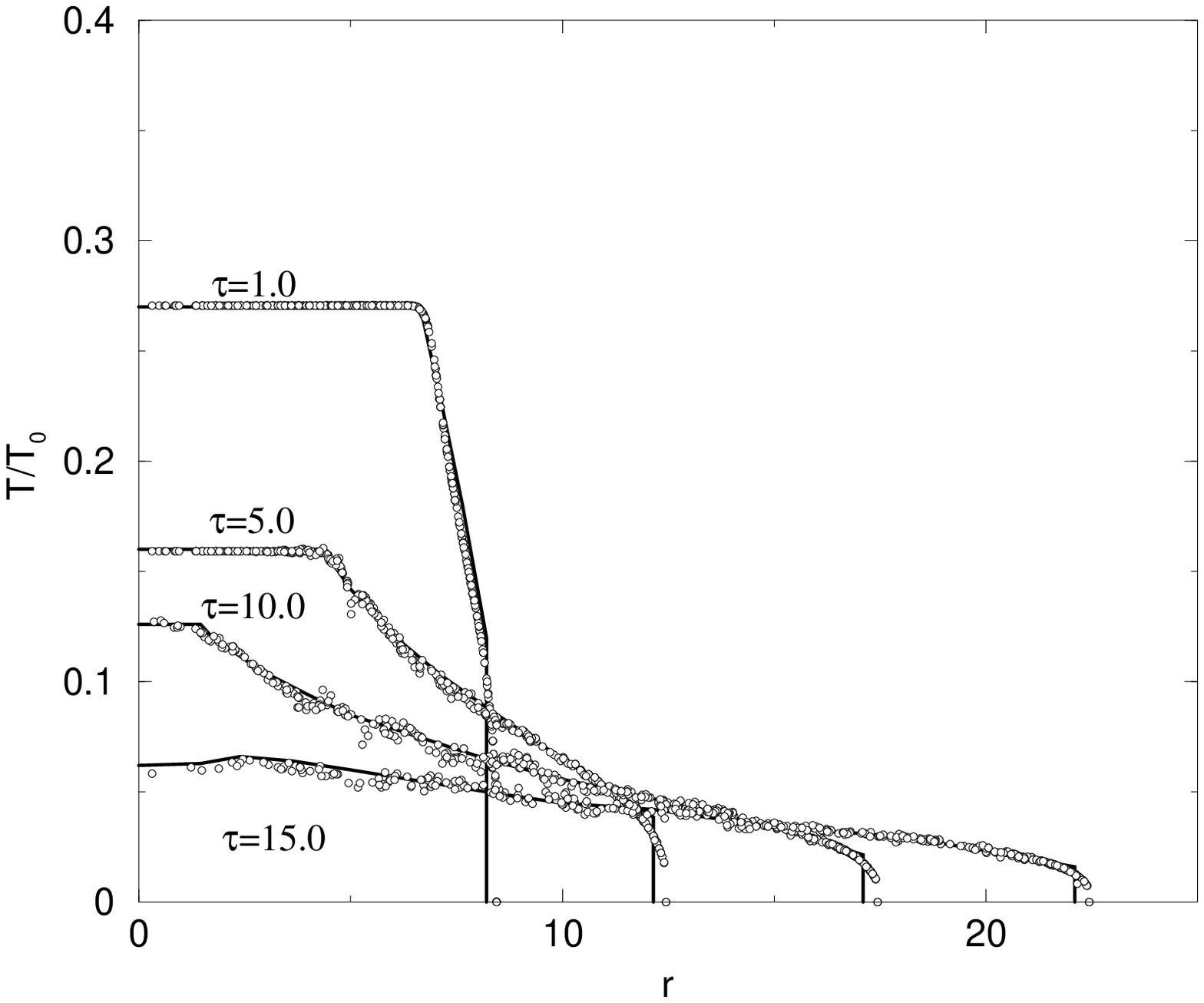}}
\vspace*{0.0cm} 
\caption{Numerical results by SPH (open circles) for the 
problem of longitudinally scaling hydrodynamics using 
the Landau-type initial conditions in the transverse directions. 
The solid line has been  obtained numerically 
by Hama and Pottag \cite{Pottag84}.}} 
\end{figure} 
\clearpage 
\begin{figure}[tbh]
{\vspace*{+0.7cm} {\hspace*{1.5cm} \epsfysize=11.0cm %
\epsfbox{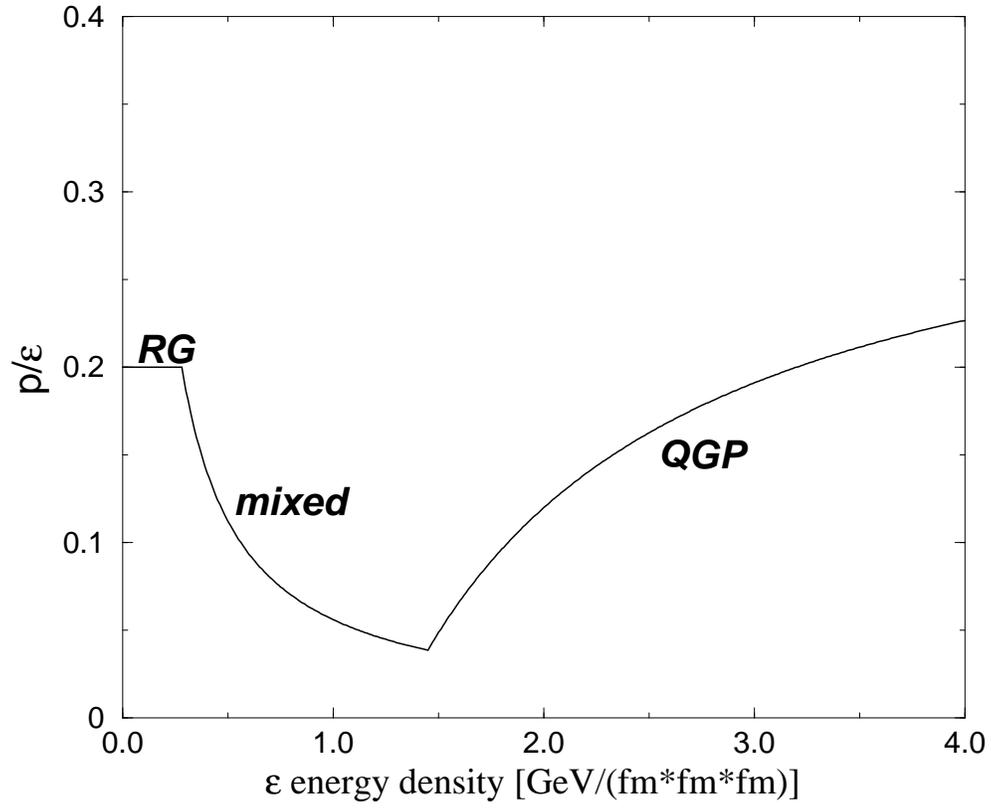}}}
\vspace*{0.0cm}
\caption{Equation of state which contains the 
QGP phase transition\cite{Hung95}. For resonance gas 
case, $c_s^2$=0.2(constant) is used.}
\end{figure}
\clearpage 
\begin{figure}[tbh]
{\vspace*{-1.0cm} \hspace*{-1.0cm} \epsfbox{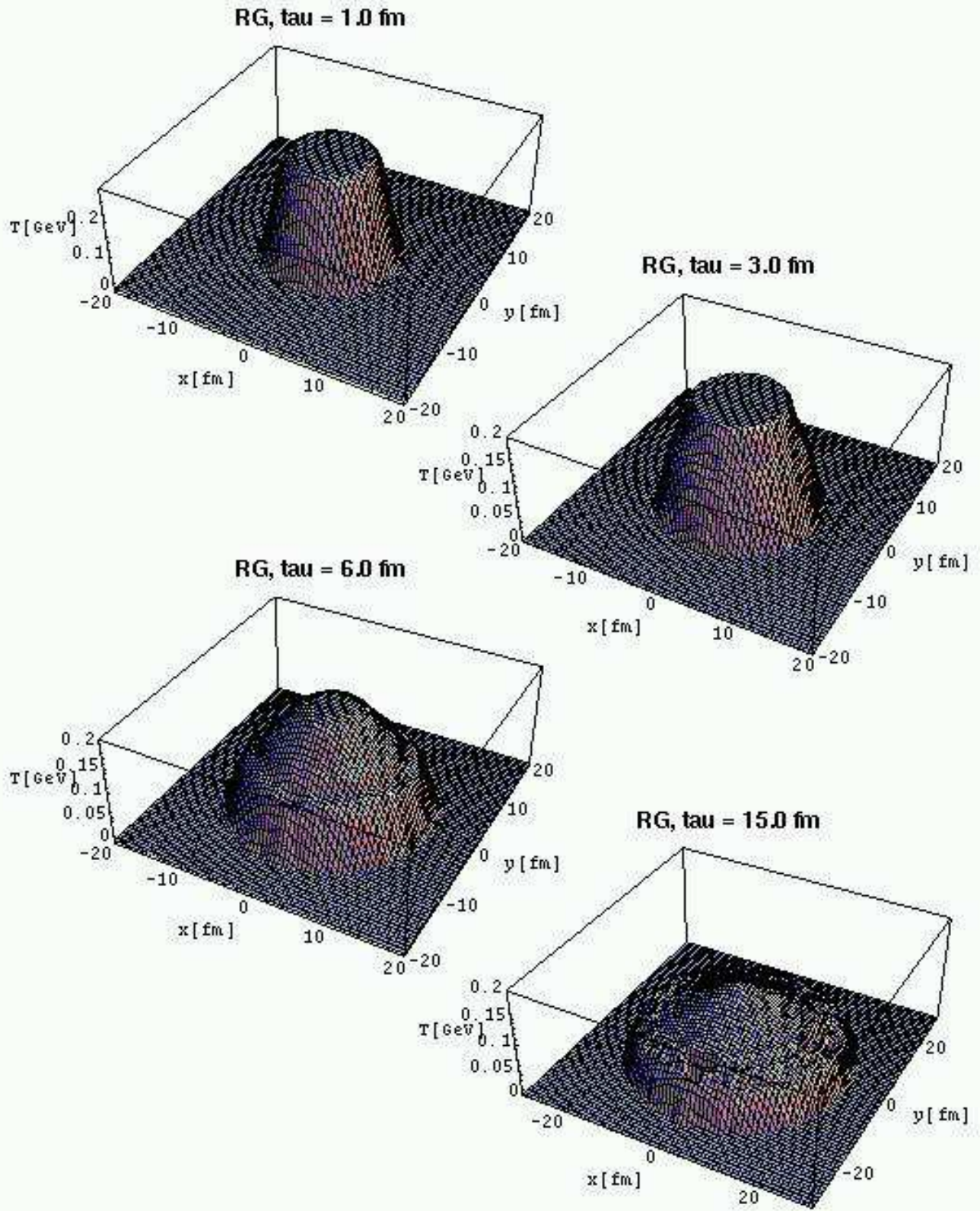}}\\

\caption{Hydrodynamical evolution
(temperature $T~[GeV]$ at $z=0 fm$ plane) using RG EoS. 
The smoothing scale parameter $h$ used in $x$-$y$ space
is $1.0~fm$ and in $\eta$ space $1.0$, respectively.} 
\end{figure}
\clearpage 
\begin{figure}[tbh] 
{\vspace*{-1.0cm} \hspace*{-1.0cm} \epsfbox{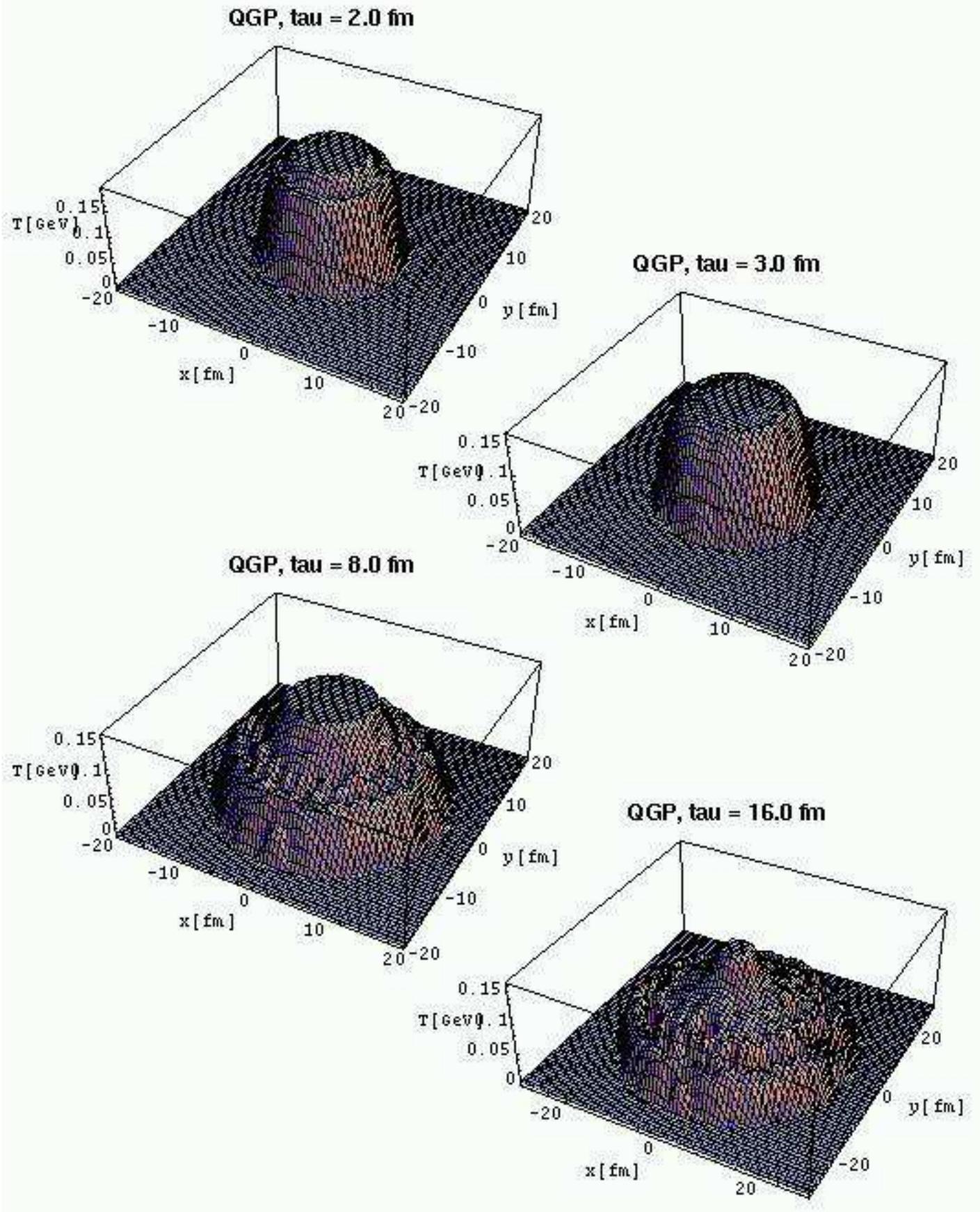}}\\ 

\vspace*{0cm} 

\caption{Hydrodynamical evolution 
(temperature $T~[GeV]$ at $z=0 fm$ plane) using QGP EoS. 
At~$\tau=2.0~fm$, we can observe 
QGP phase (above $T>T_c=0.16~GeV$) and mixed phase($T=T_c$) clearly.
At around $\tau\sim 3~fm$, QGP phase almost disappear. Mixed phase 
survives up to about 25~$fm$.
}
\end{figure} 
\clearpage 
\begin{figure}[tbh] 
{\vspace*{-1.0cm} \hspace*{-1.0cm} \epsfbox{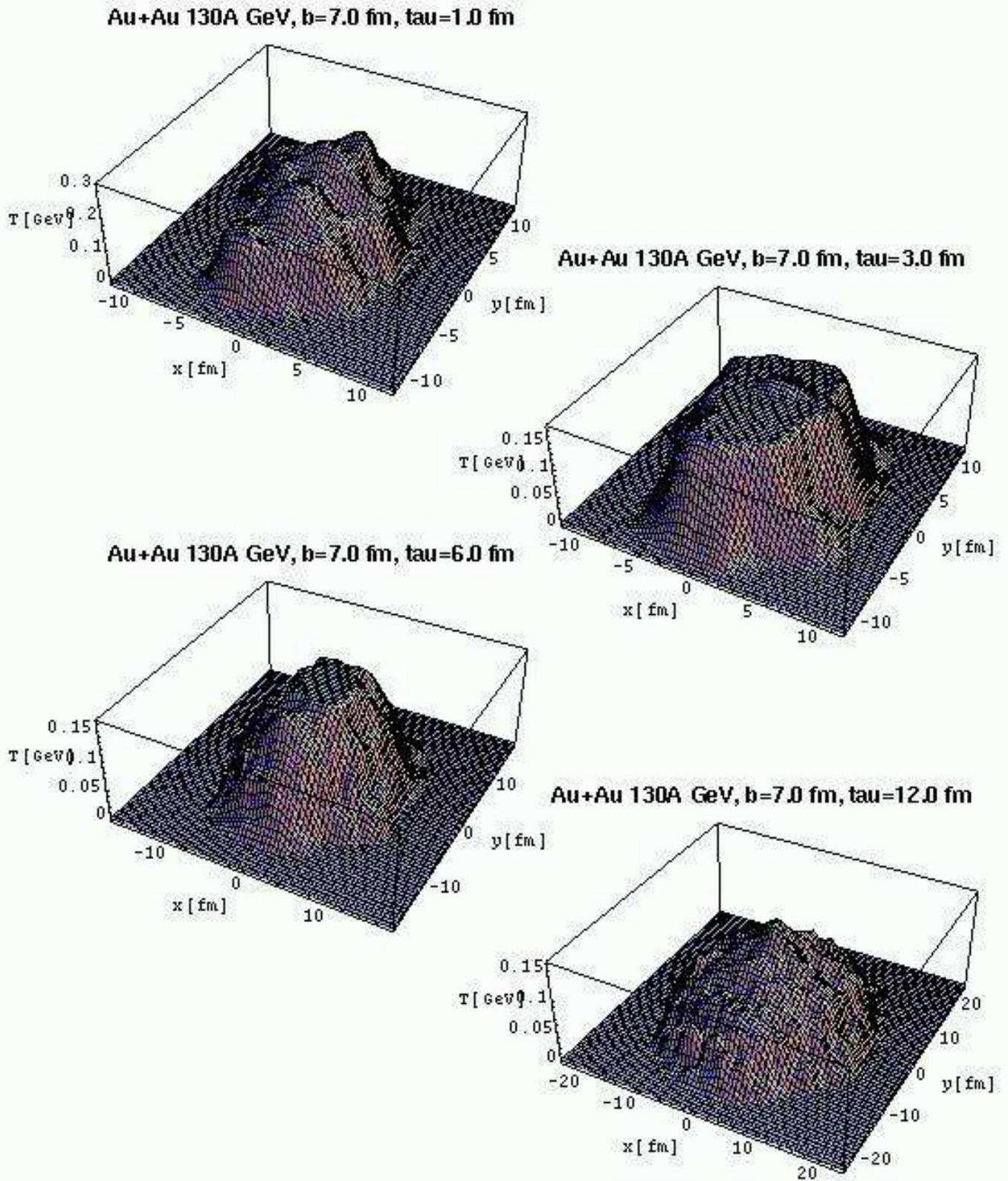}}\\

\vspace*{-0cm} 

\caption{
Hydrodynamical evolution (temperature $T~[GeV]$) of 
a typical event in $Au+Au$ collision at energy $\sqrt{s}=130A~GeV$ 
for the  impact parameter $b=7.0~fm$ (at $z=0~fm$, 
impact parameter direction is parallel to $x$ direction). 
The initial conditions are produced by NeXus\cite{NeXus}. 
The EoS used is QGP one. 
}
\end{figure}
\begin{figure}[tbh]
{\vspace*{+0.7cm} {\hspace*{1.5cm} \epsfysize=10.0cm %
\epsfbox{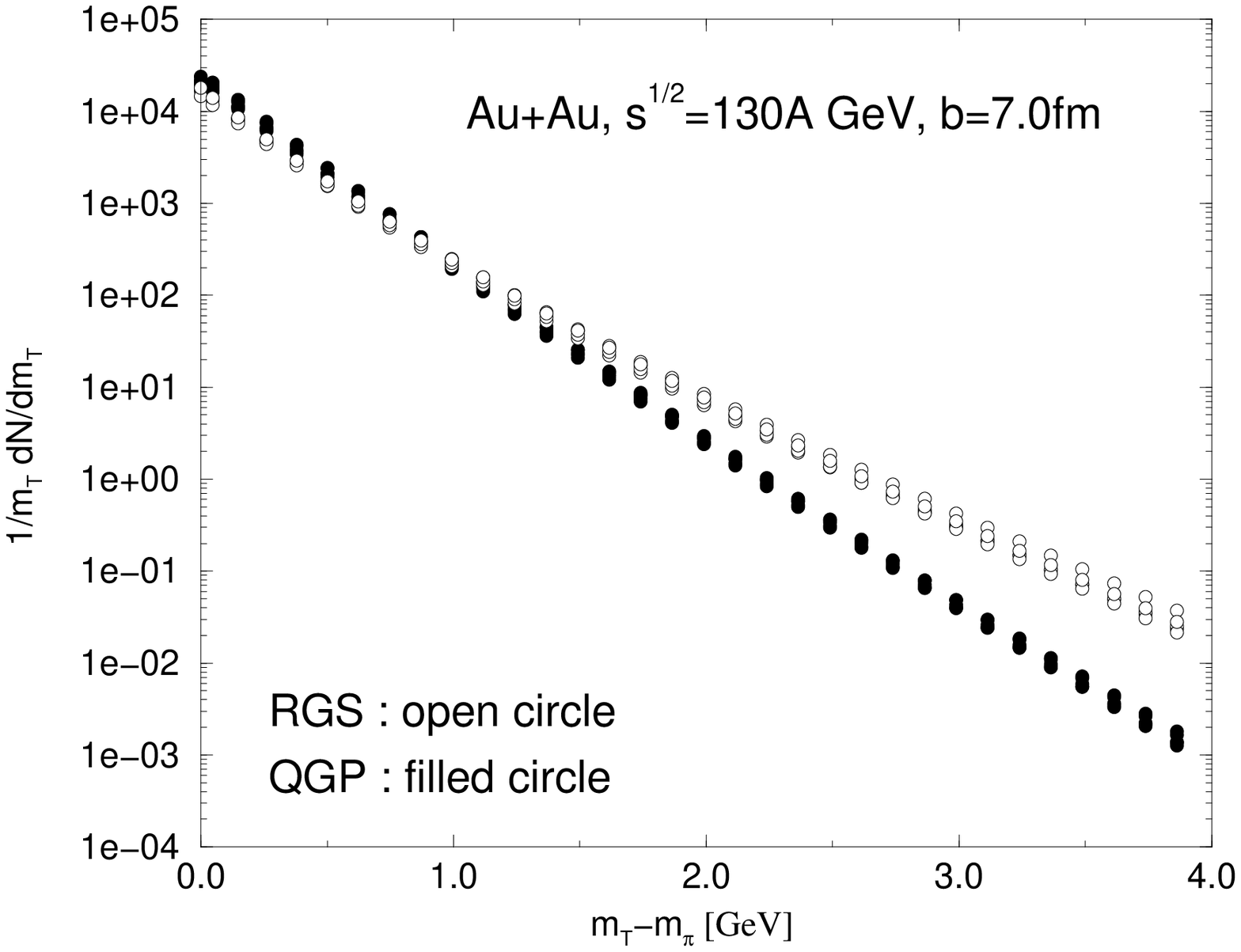}}}
\vspace*{0.0cm} 
\caption{$\mt$-spectra in $Au+Au$ collisions at $\sqrt{s}=130A~GeV$, 
$b=7.0~fm$, for 5 events (open circles for RG and filled circles for QGP).
} 
{\vspace*{+0.7cm} {\hspace*{1.5cm} \epsfysize=10.0cm %
\epsfbox{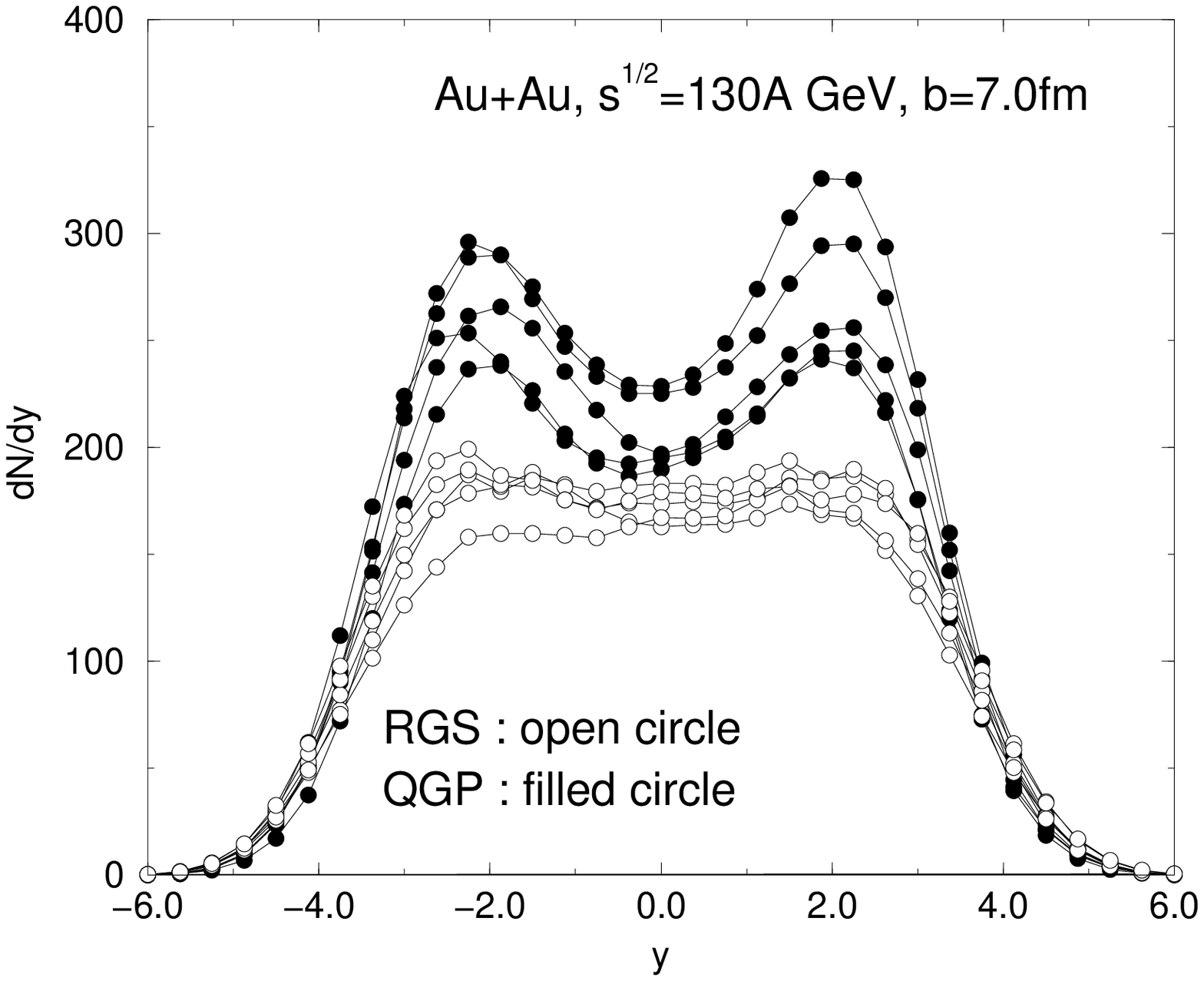}}}
\vspace*{0.0cm} 
\caption{$dN/dy$ in $Au+Au$ collisions at $\sqrt{s}=130A~GeV$, 
$b=7.0~fm$, for 5 events (open circles for RG and filled circles for QGP).
} 
\end{figure} 
\clearpage 
\begin{figure}[tbh]
{\vspace*{+0.7cm} {\hspace*{1.5cm} \epsfysize=10.0cm %
\epsfbox{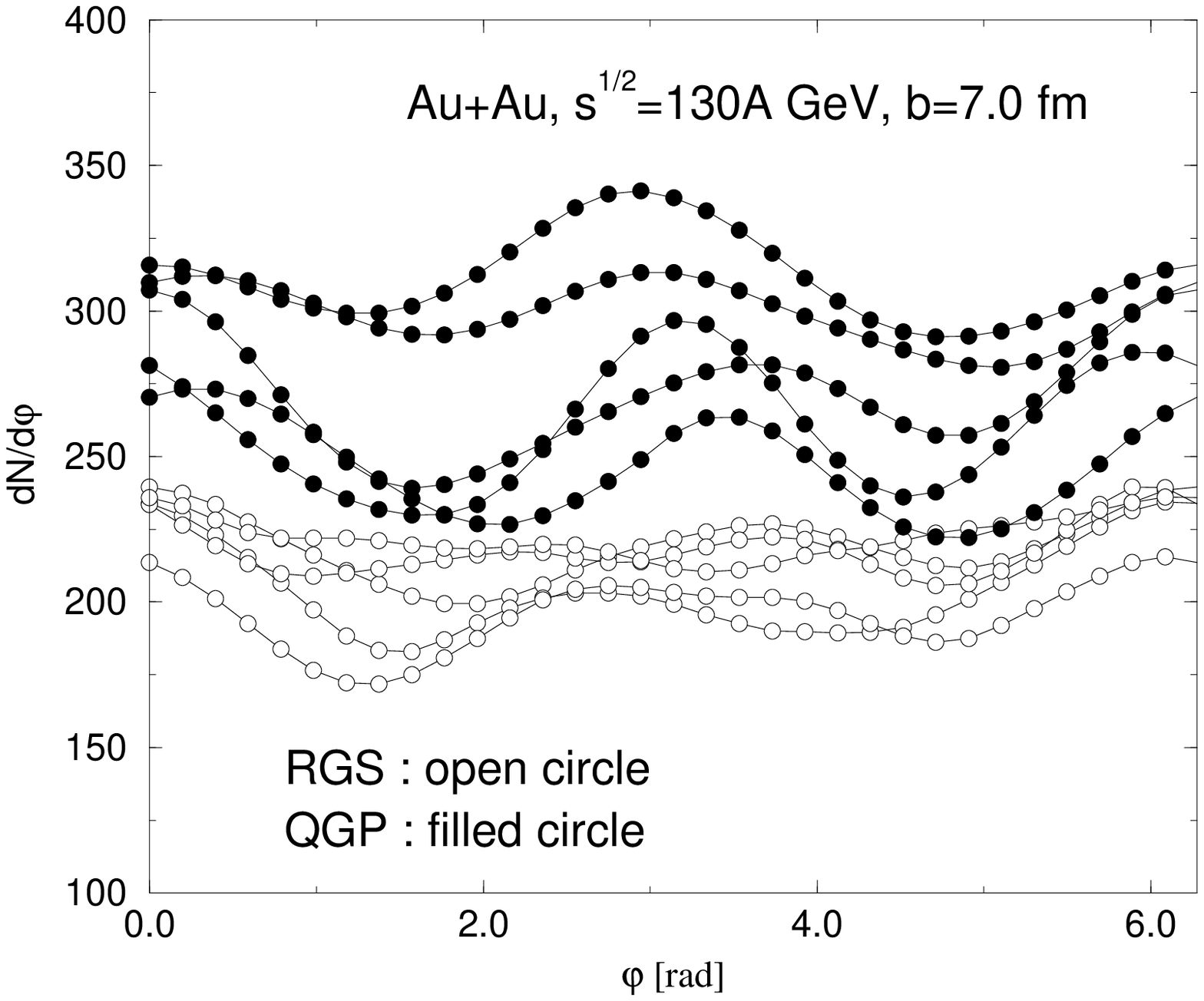}}}
\vspace*{0.0cm} 
\caption{$dN/d\varphi$ in $Au+Au$ collisions at $\sqrt{s}=130A~GeV$, 
$b=7.0~fm$, for 5 events (open circles for RG and filled circles for QGP).
} 
\end{figure} 

\clearpage 
\begin{figure}[tbh]
{\vspace*{+0.7cm} {\hspace*{1.5cm} \epsfysize=10.0cm %
\epsfbox{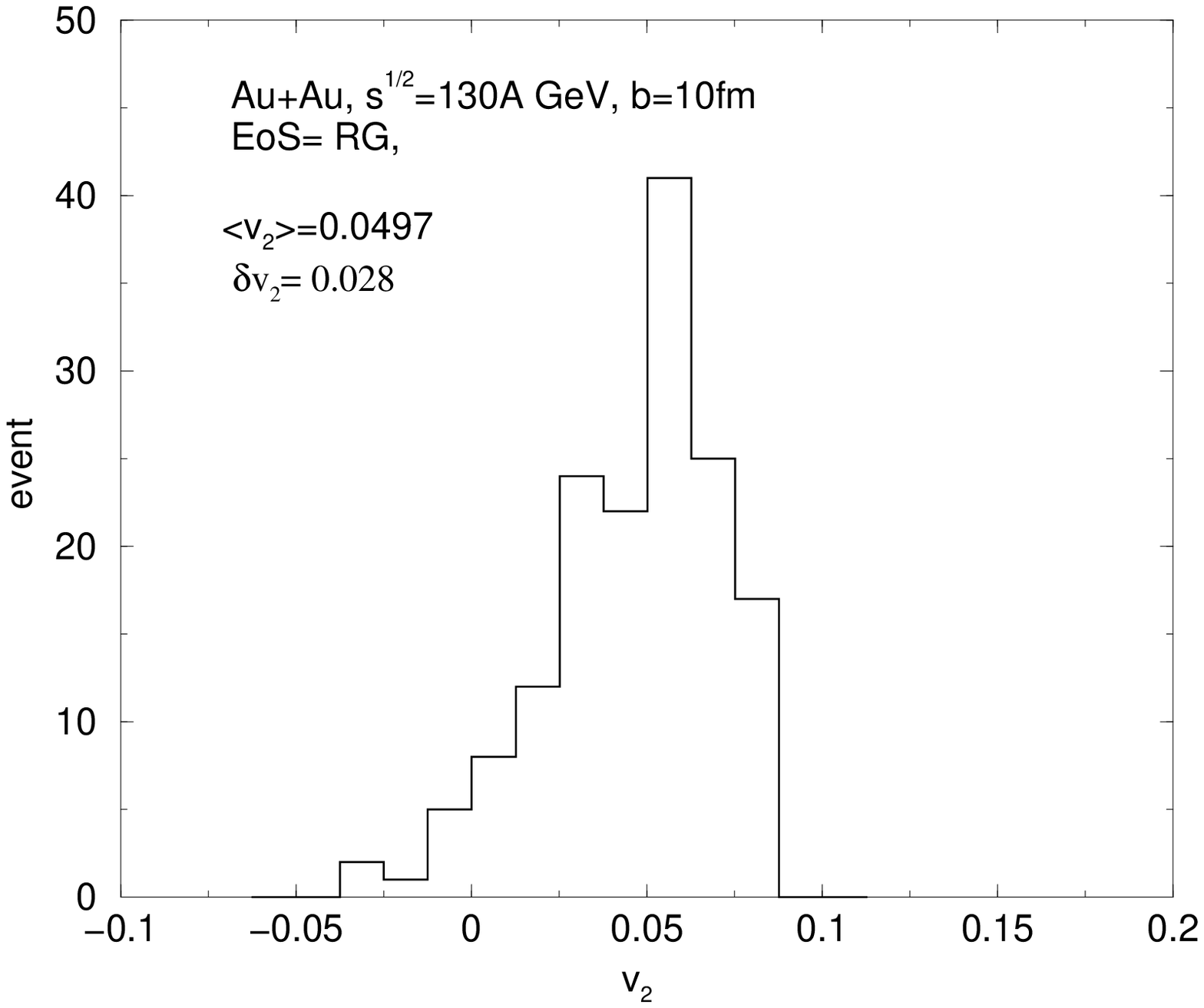}}}
\vspace*{0.0cm} 
\caption{The distribution of elliptic-flow coefficients $v_2$ for 
$Au+Au$ collisions at energy $\sqrt{s}=130A~GeV$($b=10~fm$). 
RG EoS is used. 
} 
{\vspace*{+0.7cm} {\hspace*{1.5cm} \epsfysize=10.0cm %
\epsfbox{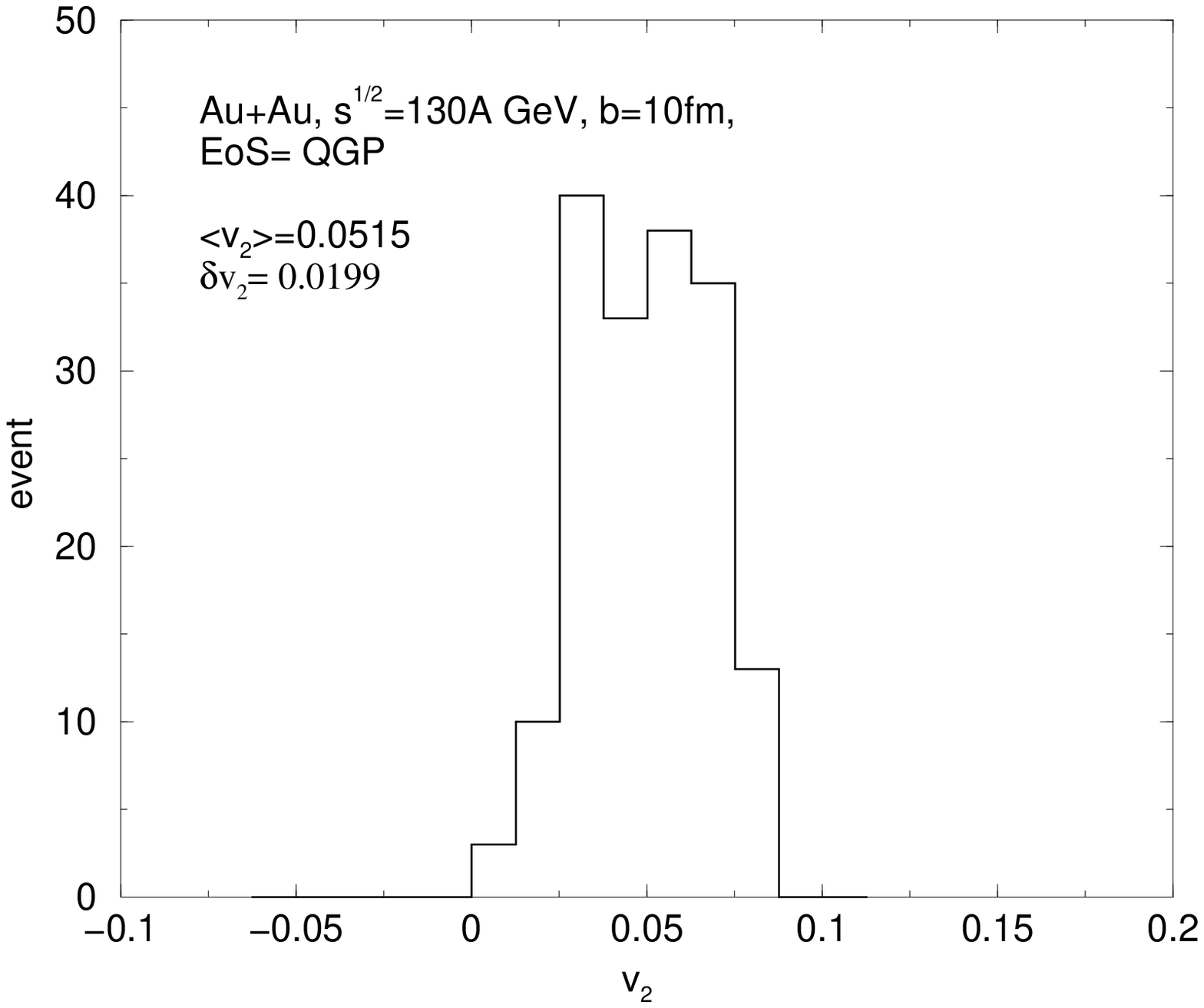}}}
\vspace*{0.0cm} 
\caption{The distribution of elliptic-flow coefficients $v_2$ for 
$Au+Au$ collisions at energy $\sqrt{s}=130A~GeV$($b=10~fm$). 
QGP EoS is used.
} 
\end{figure} 
\clearpage 
\begin{figure}[tbh]
{\vspace*{+0.7cm} {\hspace*{1.5cm} \epsfysize=10.0cm %
\epsfbox{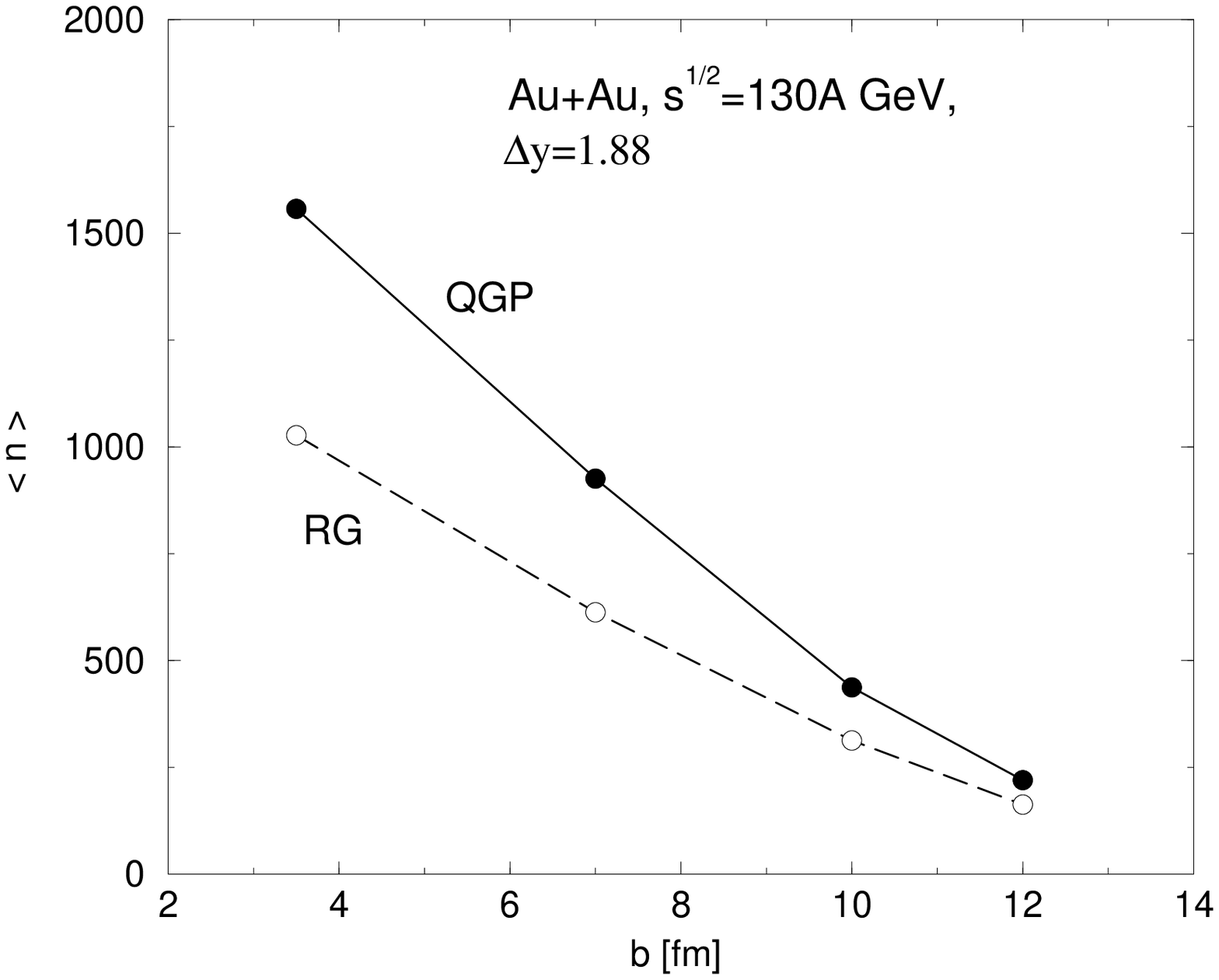}}}
\vspace*{0.0cm} 
\caption{
EoS dependence of $\la n(y,\Delta y) \ra$ ($y=0,\Delta y= 1.875$) 
as function of impact parameter $b$.} 
{\vspace*{+0.7cm} {\hspace*{1.5cm} \epsfysize=10.0cm %
\epsfbox{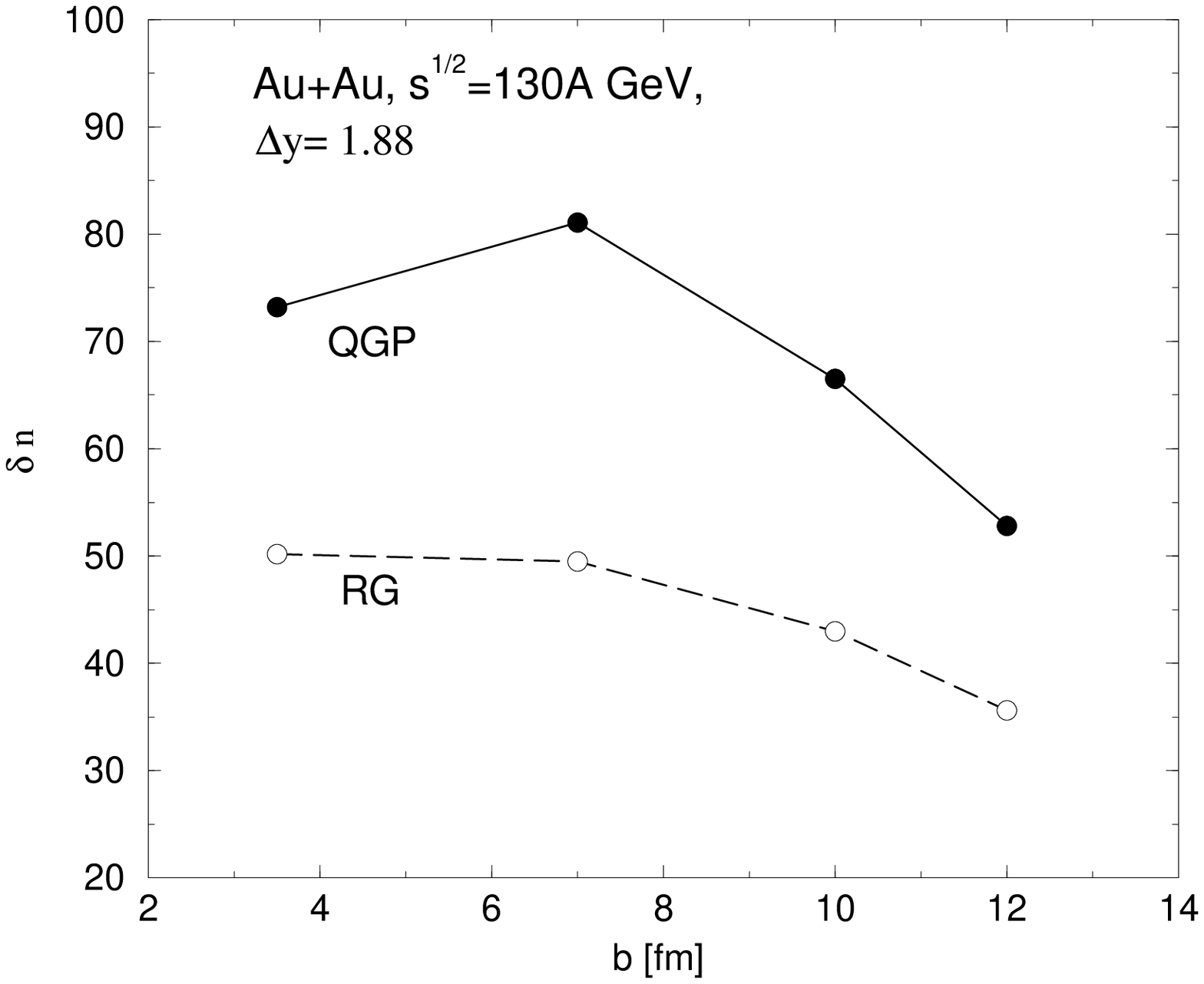}}}
\vspace*{0.0cm} 
\caption{EoS dependence of $\delta n(y,\Delta y)$ ($y=0,\Delta y= 1.875$) 
as function of impact parameter $b$.
} 
\end{figure} 
\begin{figure}[tbh]
{\vspace*{+0.7cm} {\hspace*{1.5cm} \epsfysize=10.0cm %
\epsfbox{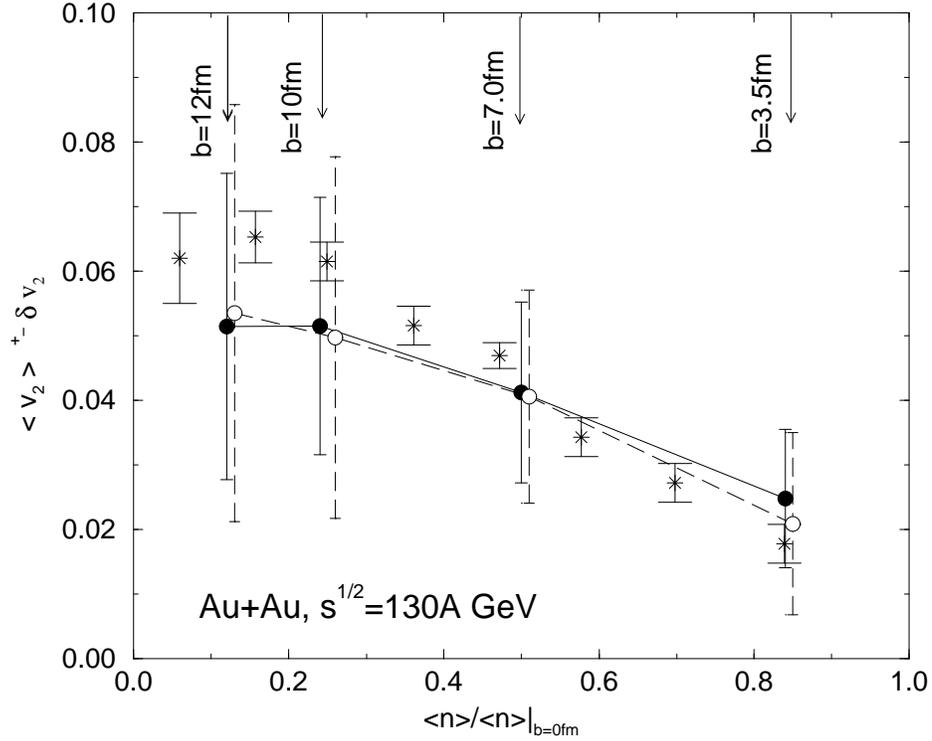}}}
\vspace*{0.0cm} 
\caption{Comparison of our result for 
$\la v_2 \ra$ with STAR collaboration data\cite{STAR00}. 
Open circles for RG and filled circles for QGP EoS. 
Whereas in the data points error bars are shown, 
in our estimates we have put $\delta v_2$.} 
\end{figure} 
\clearpage 
\begin{table}[b] 
\newcommand{\lw}[1]{\smash{\lower2.0ex\hbox{#1}}} 
\renewcommand{\arraystretch}{1.2} 
\begin{center} 
\begin{tabular}{|c|c|c|r|r|}
\hline
\quad $b~[fm]$ \quad& \quad EoS \quad\quad & event &
$\la v_2\ra$&$\delta v_2$\\ \hline \hline 
\lw{3.5}   & RG  & 68& 
0.0219 & 0.0141 \\
\cline{2-5}& QGP & 558& 
0.0242 & 0.0107 \\
\hline\hline 
\lw{7.0}   & RG  & 68& 
0.0406 & 0.0165 \\
\cline{2-5}& QGP & 72& 
0.0412 & 0.0140 \\
\hline\hline 
\lw{10.0}  & RG  &166& 
0.0497 & 0.0280 \\
\cline{2-5}& QGP &180& 
0.0515 & 0.0199 \\
\hline\hline 
\lw{12.0}  & RG  & 95& 
0.0535 & 0.0323 \\
\cline{2-5}& QGP &119& 
0.0514 & 0.0237 \\
\hline 
  \end{tabular}
  \end{center} 
  \caption{The average value and dispersions of the 
directed$(v_1)$ and elliptic$(v_2)$ flow coefficients 
for $Au+Au$ collisions at energy $\sqrt{s}=130A~GeV$.} 
 \end{table} 
\begin{table}[b] 
\newcommand{\lw}[1]{\smash{\lower2.0ex\hbox{#1}}} 
\renewcommand{\arraystretch}{1.2} 
\begin{center} 
\begin{tabular}{|c|c|c|c|c|}
\hline
\quad $b~[fm]$ \quad& \quad EoS \quad\quad & event 
&\quad\quad$\tilde{T}$ &\quad\quad$\delta\tilde{T} $ \\ \hline \hline 
\lw{3.5}   & RG  & 68& 0.229 & 0.0024 \\ 
\cline{2-5}& QGP & 55& 0.214 & 0.0014 \\
\hline\hline 
\lw{7.0}   & RG  & 55& 0.231 & 0.0032 \\ 
\cline{2-5}& QGP & 58& 0.215 & 0.0021 \\
\hline\hline 
\lw{10.0}  & RG  & 90& 0.233 & 0.0041 \\
\cline{2-5}& QGP &119& 0.214 & 0.0026 \\
\hline\hline 
\lw{12.0}  & RG  & 79& 0.234 & 0.0047 \\
\cline{2-5}& QGP &100& 0.213 & 0.0033 \\ 
\hline 
  \end{tabular}
  \end{center} 
  \caption{The slope parameter $\tilde{T}$ and its dispersion 
$\delta \tilde{T}$.} 
 \end{table} 

 \begin{table}[b] 
\newcommand{\lw}[1]{\smash{\lower2.0ex\hbox{#1}}} 
\renewcommand{\arraystretch}{1.2} 
\begin{center} 
\begin{tabular}{|c|c|c|r|r|r|r|r|r|}
\hline
\lw{$b~[fm]$}& \quad \lw{EoS} \quad & \lw{event} 
&\multicolumn{3}{c|}{$\Delta y$=1.875} &
\multicolumn{3}{c|}{$\Delta y$=3.00}\\ 
\cline{4-9} & & & \quad $\la n \ra $\quad & 
\quad $\delta n $ & $\delta n $/$\la n \ra $ 
&\quad $\la n  \ra $ &\quad$\delta n $\quad 
& $\delta n $/$\la n \ra $
\\ \hline\hline 
\lw{3.5}    & RG  & 68 & 1026.5& 50.2 & 0.049     &1619.2& 73.2   &0.045\\
\cline{2-9} & QGP & 55 & 1557.5& 73.2 & 0.047     &2548.4& 114.4  &0.045\\
\hline \hline 
\lw{7.0}    & RG  & 55 & 613.3 & 49.5 & 0.081     &977.4 & 71.6   &0.073\\
\cline{2-9} & QGP & 58 & 926.1 & 81.1 & 0.087     &1530.5& 123.7  &0.081\\
\hline \hline 
\lw{10.0}   & RG  &166 & 312.8 & 43.0 & 0.137     &506.1 &  65.8  &0.130\\
\cline{2-9} & QGP &180 & 437.5 & 66.5 & 0.151     &740.7 & 103.9  &0.140\\
\hline \hline 
\lw{12.0}   & RG  & 79 & 162.8 & 35.6 & 0.219     & 268.9 & 56.2  & 0.209\\
\cline{2-9} & QGP &100 & 220.1 & 52.8 & 0.240     & 379.8 & 85.2  & 0.224\\
\hline 
  \end{tabular}
  \end{center} 
  \caption{EoS dependence of the multiplicity fluctuation in two
    different $\Delta y$ around $y=0$, as function of the 
impact parameter $b$.} 
\end{table}
\end{document}